\documentclass[conference]{IEEEtran}

\usepackage[dvips]{color}
\usepackage{epsf}
\usepackage{times}
\usepackage{epsfig}
\usepackage{graphicx}
\usepackage{float} 
\usepackage{amsmath}
\usepackage{amssymb}
\usepackage{amsxtra}
\usepackage{amsthm}
\usepackage{bbm}

\usepackage{here}
\usepackage{rawfonts}
\usepackage{times}
\usepackage{url}
\usepackage{cite}
\usepackage{comment}
\usepackage[utf8]{inputenc}
\usepackage{caption}
\usepackage{subcaption}
\usepackage[normalem]{ulem}

\usepackage{pstricks}
\usepackage{algorithm}
\usepackage{algpseudocode}
\usepackage{lipsum}

\IEEEoverridecommandlockouts 

\begin{document}
\vspace{-0.0cm}
\title{\huge
	Privacy Preserving Semantic Communications Using Vision Language Models: A Segmentation and Generation Approach
}\vspace{-0.0cm}	
\author{ 
	\IEEEauthorblockN{ 
	Haoran Chang$^\ast$, Mingzhe Chen$^\dagger$$^\S$, Huaxia Wang$^\ast$, and Qianqian Zhang$^\ast$} 
	
	\IEEEauthorblockA{\small
		$^\ast$Department of Electrical and Computer Engineering, Rowan University,  NJ, USA,
		Emails: \url{{changh35,wanghu,zhangqia}}@rowan.edu \\
        $^\dagger$Department of Electrical and Computer Engineering, University of Miami,   FL, USA,
        Emails: \url{{mingzhe.chen}}@miami.edu \\
        $^\S$Frost Institute for Data Science and Computing, University of Miami,   FL,  USA
		%\thanks{This work was supported in part by the U.S. National Science Foundation under Grant ECCS-2434054. }
        \vspace{-0.2cm}				
	}
}
\maketitle

\begin{abstract} 

Semantic communication has emerged as a promising paradigm for next-generation wireless systems, improving the communication efficiency by transmitting high-level semantic features. 
However, reliance on unimodal representations can degrade reconstruction under poor channel conditions, 
and privacy concerns of the semantic information attack also gain increasing attention. 
In this work, a privacy-preserving semantic communication framework is proposed to protect sensitive content of the image data. 
Leveraging a vision-language model (VLM), the proposed framework identifies and removes private-content regions from input images prior to transmission. 
A shared privacy database enables semantic alignment between the transmitter and receiver to ensure consistent identification of sensitive entities. 
At the receiver, a generative module reconstructs the masked regions using learned semantic priors and conditioned on the received text embedding.  
Simulation results show that generalizes well to unseen image processing tasks, improves reconstruction quality at the authorized receiver by over $10\%$ using text embedding, and reduces identity leakage to the eavesdropper by more than $50\%$.

\end{abstract}  
 
%\IEEEpeerreviewmaketitle

\vspace{-0.1cm}
\section{Introduction}

Semantic communication has emerged as a promising paradigm for next-generation communication systems. 
Unlike traditional approaches that focus on the accurate delivery of bit sequences, semantic communication aims to convey the underlying  meaning of the communication data \cite{xie2021deep}. 
By leveraging advances in natural language processing and computer vision, this approach  enables context-aware data exchange to improve transmission efficiency.  
Generally, the transmitter of semantic communication extracts semantic features from the source data prior to conventional bit-level and channel-level encoding, and thus, reduces the amount of data transmitted over the communication system. 
The receiver then reconstructs the information to preserve semantic fidelity, and ensures the intended meaning to be accurately conveyed.   
As a result, semantic communication offers the potential to support a wide range of emerging 6G applications, including augmented/virtual reality (AR/VR) and autonomous systems.

Existing works in \cite{xie2021deep,zhang2024performance,yoo2022real,hu2023robust} have explored semantic communication across various applications. 
In \cite{xie2021deep},  a deep learning-based framework was proposed that jointly trains semantic and channel encoders/decoders to extract essential features and ensure robust transmission over physical channels.  
To enable accurate extraction and reconstruction of textual messages, \cite{zhang2024performance} incorporated a shared knowledge base to align prior semantic information between transmitter and receiver, thus facilitating semantic interpretation.   
For image transmissions, deep neural networks (NNs) such as variational autoencoder (VAE) \cite{hu2023robust} and vision transformer \cite{yoo2022real} have been employed to  encode visual content by feature extraction and image reconstruction.    
However, these methods rely solely on single-modal representations and lack explicit semantic reasoning, which cannot support cross-modal interpretability or high-level content understanding.   
To address these limitations, recent works in \cite{cicchetti2024language} and \cite{zhao2024lamosc} leveraged  the vision-language model (VLM)  to enhance the semantic extraction and representation through multi-modalities processing. 
In these approaches, the transmitter uses a VLM to convert input images into textual descriptions and extract latent embeddings that retain perceptual and semantic details. At the receiver, the image is regenerated using both the textual description and latent vectors to achieving high reconstruction quality.  
Despite these advantages, VLM-based semantic communication introduces new security concerns, where the transformation of images into textual and latent representations increases the risk of semantic leakage, as sensitive content may be exposed through intermediate features, even after compression.

To address the challenge of privacy leakage in semantic communication, this paper proposes a novel VLM-based framework to identify and remove sensitive content prior to communication. 
In scenarios where adversaries intercept semantic data, the framework leverages shared semantic knowledge, which is established through a pre-defined privacy image dataset, to align the transmitter and receiver on what constitutes sensitive information. 
At the transmitter, a semantic segmentation module detects and masks privacy regions in the image. 
The receiver then reconstructs the masked content using a generative model guided by the shared semantic priors and the received text embedding.  
Simulation results show that the proposed method enhances the image transmission quality for authorized users, significantly reduces the identity leakage to unauthorized parties, and exhibits strong generalization to unseen image processing tasks.   
To the best of our knowledge, this is the first work to apply VLMs for privacy-preserving semantic communication to enhance the security of future wireless networks.

The rest of this paper is organized as follows.  Section \ref{sysModel_proFormulation} introduces  the system model and problem formulation of semantic communication. Section \ref{solution} presents the proposed privacy-preserving solution. Simulation results are shown in Section \ref{simulation}, and conclusions are drawn in Section \ref{conclusion}.

\section{System Model and Problem Formulation}\label{sysModel_proFormulation}

\begin{figure}[t] %\vspace{0.1cm}
    \centering
    \includegraphics[width=8cm]{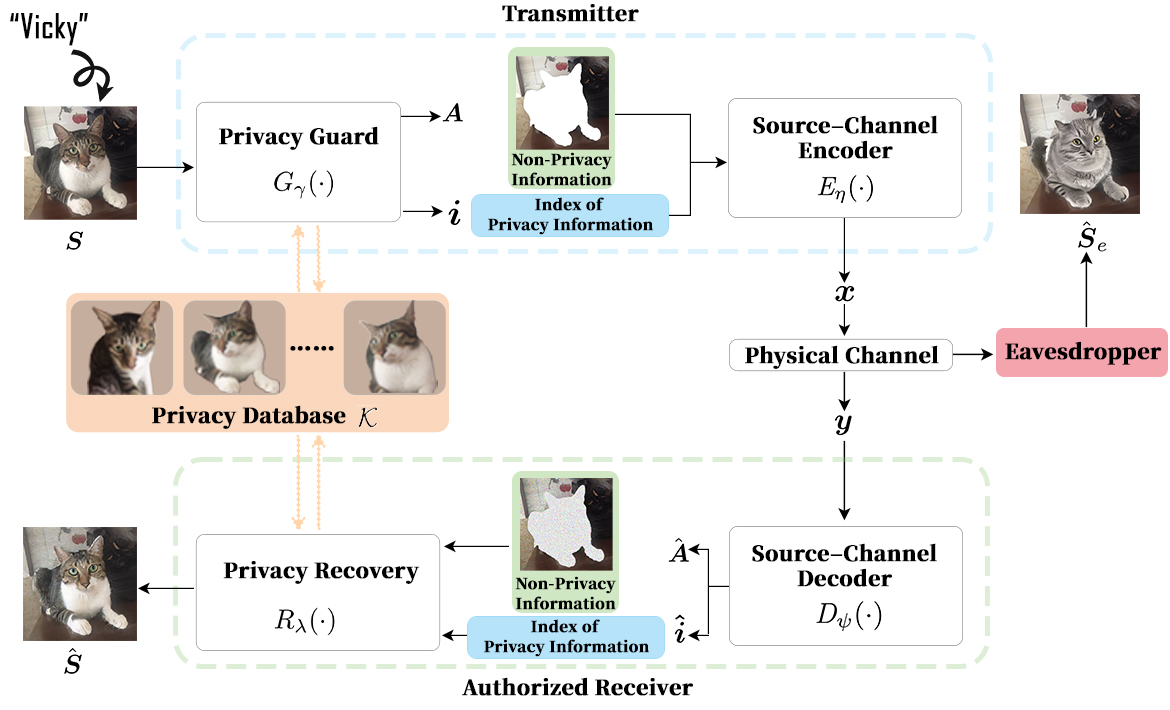}\vspace{-0.2cm}
    \caption{Privacy-Preserving System Model Overview} \vspace{-0.6cm}
    \label{fig:system_model}   
\end{figure}

%\vspace{-0.1cm}
\subsection{System Model}  

We consider a wireless system where a transmitter sends image data to an authorized receiver, while unauthorized entities may attempt to intercept and extract sensitive information.  
To safeguard privacy, a semantic protection framework, comprising a privacy guard module, a privacy database, and a recovery module, is proposed, as shown in Fig. \ref{fig:system_model}. 
The system removes sensitive content from the image prior to transmission to ensure no privacy-related information is present in the transmission.  
Privacy definitions are established based on shared knowledge between the transmitter and the authorized receiver, and stored in a privacy database $\mathcal{K}$  constructed during initialization. 
For example, if a cat named Vicky is marked as sensitive, its visual data is excluded from the transmission.
The authorized receiver, using the shared database, semantically reconstructs the full image, while an eavesdropper without access to this knowledge cannot infer the omitted content.

At the transmitter, the input image $\boldsymbol{S} \in \mathbb{R}^{C \times H \times W}$ is first processed by the privacy guard module $G_{\gamma}(\cdot)$ to  identify and remove the sensitive content \cite{kirillov2023segment}, 
where $\mathit{C}$, $\mathit{H}$, and $\mathit{W}$ denote the channel, height and width of the image, respectively, and  $\gamma$ is the trainable parameter. 
The output of this process is: 
\begin{equation}
    (\boldsymbol{A}, \boldsymbol{i}) = G_\gamma(\boldsymbol{S}| \mathcal{K} ),
\end{equation} 
where $\boldsymbol{A} \in \mathbb{R}^{C \times H \times W}$ is the processed image with private content removed, and $\boldsymbol{i} \in \mathbb{R}^d$ is the text embedding representing the index of identified privacy content.  
In our model, $\boldsymbol{i}$ can be derived from a set of predefined prompt (e.g., Remove the cat Vicky), and then, transferred into a numerical vector that captures the semantic meaning of a textual description in a format that NNs can process. 
In the example of Fig. \ref{fig:system_model}, $\boldsymbol{i}$ represents the semantic index corresponding to Vicky and $\boldsymbol{A}$ denotes the masked image with Vicky’s region removed.

After semantic segmentation,  $\boldsymbol{A}$ and $\boldsymbol{i}$ are encoded by $E_{\eta}(\cdot)$ into a transmit signal $\boldsymbol{x}$, i.e.,
\begin{equation}\vspace{-0.1cm}
    \boldsymbol{x} = E_\eta(\boldsymbol{A}, \boldsymbol{i}), 
\end{equation}
where $\eta$ is the trainable parameter. The signal $\boldsymbol{x}$ is then transmitted over a wireless channel $\boldsymbol{H}$ with additive white Gaussian noise (AWGN), and the authorized user receives 
\begin{equation}
    {\boldsymbol{y}} = \boldsymbol{H} \boldsymbol{x} + \boldsymbol{n}, 
\end{equation}
where  $\boldsymbol{n} \sim \mathcal{CN}(0, \sigma_n^2 \mathbf{I})$  is the receiver noise. 

Upon receiving the signal, the authorized receiver applies the decoder $D_{\psi}(\cdot)$ with parameter $\psi$ to obtain the masked image $\hat{\boldsymbol{A}}\in \mathbb{R}^{C \times H \times W}$ and the privacy index $\hat{\boldsymbol{i}}\in \mathbb{R}^d$, via
\begin{equation}
    (\hat{\boldsymbol{A}}, \hat{\boldsymbol{i}}) = D_{\psi}( \boldsymbol{y} ).
\end{equation}

The decoded messages are then passed through the privacy recovery module $R_{\lambda}(\cdot)$, based on a VLM with  the model parameter $\lambda$, to reconstruct the complete image as 
\begin{equation}
\hat{\boldsymbol{S}} = R_\lambda(\hat{\boldsymbol{A}}, \hat{\boldsymbol{i}}| \mathcal{K}).
\end{equation} 

Meanwhile, the unauthorized receiver may intercept the transmitted signal. 
Here, we assume a powerful eavesdropper equipped with its own decoding $D_e$ and generation module $R_e$ that are similar to these of the authorized receiver. 
The recovered image at the eavesdropper is 
\begin{equation}
    \hat{\boldsymbol{S}}_e =R_e(D_e(\boldsymbol{y}_e | \boldsymbol{x} (\gamma, \eta) )).
\end{equation}

However, without access to the privacy database $\mathcal{K}$, the eavesdropper can only partially recover the image with the privacy content missing or mis-represented. In the example shown in Fig.1, the eavesdropper may infer that the masked image is a cat,  but it cannot clearly identify its color or fur pattern, thus preserving the identity information about Vicky.

\subsection{Performance Metrics}  

The goal of privacy-preserving semantic communication is to minimize the distortion $\mathcal{L}_{a}$ between the input $\boldsymbol{S}$ and the reconstructed image $\hat{\boldsymbol{S}}$ at the authorized receiver, as well as to minimize the privacy leakage  ${f}_{e}$ at the eavesdropper. 
Specifically, the distortion loss is defined as the normalized reconstruction error: 
\begin{equation}
    \mathcal{L}_a (\gamma, \eta, \psi,\lambda) =  \frac{|| \boldsymbol{S} - \hat{\boldsymbol{S}}(\gamma, \eta, \psi,\lambda) ||^2}{255 \cdot C \cdot H \cdot W} \in [0,1].   
\end{equation}
% which quantifies . 
To quantify the privacy leakage at the eavesdropper, the generated image $ \boldsymbol{\hat{S}}_e$ will be evaluated by the privacy guard module via $(\boldsymbol{A}_e, \boldsymbol{i}_e) = G_\gamma(\boldsymbol{\hat{S}}_e| \mathcal{K} )$. 
The privacy leakage function is: % defined as 
\begin{equation}
f( \boldsymbol{\hat{S}}_e  (\gamma, \eta) )= 
\begin{cases}
    1,& \text{if }  \boldsymbol{i}_e = \boldsymbol{i},\\
    0,              & \text{otherwise},
\end{cases} 
\end{equation}
which equals one if the eavesdropper successfully identifies the masked sensitive content, and zero otherwise. 

\subsection{Problem Formulation} \label{problemFormulation}
%Consequently,  
To ensure accurate image reconstruction at the authorized receiver while preventing the privacy disclosure during transmission, the problem can be formulated as: 
\begin{equation} 
    \begin{split} \label{equ_opt}
        \underset{ \gamma, \eta,\psi, \lambda  } {\min} &
        \mathbb{E} \left[ \mathcal{L}_a (\gamma, \eta, \psi,\lambda) +  f( \boldsymbol{\hat{S}}_e  (\gamma, \eta) )   \right]   \\ 
        \text{s.t.}~ &  \frac{||\boldsymbol{H} \boldsymbol{x} (\gamma, \eta)||^2}{B \sigma_n^2} \ge \tau_{\text{thre}},    \\
                     &  ||\boldsymbol{x} (\gamma, \eta)||^2 \le P_{\text{max}},  
    \end{split}
\end{equation}
where the first constraint requires the signal-to-noise ratio (SNR) of the wireless transmission to be no less than a threshold $\tau_{\text{thre}}$, and the second constraint enforces  a maximum transmission power $P_{\text{max}}$.

The  problem in (\ref{equ_opt}) is challenging to solve for three reasons. 
Firstly, the information bottleneck imposed by the wireless channel constrains the semantic communication performance. 
Thus,  channel limitations must be incorporated into the training process of the source-channel encoder and decoder. 
Secondly, the multi-modal nature of the input necessitates the use of text embeddings to guide image generation while preserving generalization capability for unseen image processing tasks. 
Thirdly, the presence of an eavesdropper complicates the design of the privacy guard. It must retain sufficient semantic information to ensure high-quality reconstruction at the authorized receiver, while effectively removing sensitive identity features to protect the privacy of the target entity.

\section{Solution} \label{solution}

This section details the design of the privacy-preserving semantic communication framework, including the privacy database, the privacy guard module, the source-channel encoder and decoder, and the privacy recovery module, with the objective of minimizing image distortion and privacy leakage.

\subsection{Privacy Database} 
The proposed semantic system operates in two stages: initialization and working. 
In the initialization stage, a privacy database is constructed based on predefined privacy entities, which are mutually agreed upon by the transmitter and the authorized receiver. 
Specifically, the database is defined  as $\mathcal{K} = \left\{ \boldsymbol{i}_k, {\mathcal{S}}_k, \boldsymbol{f}_k  \right\}_{k=1,\cdots,N}$, 
where $\boldsymbol{i}_k $ represents the index of privacy entity $k$, 
$\mathcal{S}_k =\{ \boldsymbol{\tilde{S}}_{k,m}\}_{m=1,\cdots,M} $ is a set of $M$ sample images for entity $k$ which will be used to guide the privacy guard module for privacy detection and the recovery module for image reconstruction, and  $\boldsymbol{f}_k \in \mathbb{R}^d$ is the corresponding feature vector. 
To derive $\boldsymbol{f}_k$, each image in $\mathcal{S}_k$ is first processed by the image encoder $S_{\alpha}(\cdot)$ to exact feature maps. 
Global pooling is then applied to aggregate these feature maps into a compact vector that captures the key attributes of privacy entity $k$. 
The structure and function of the encoder $S_{\alpha}(\cdot)$ will be detailed in the next section. %subsequent section of the Privacy Guard Module. 
During the following work stage, the privacy database $\mathcal{K}$ remains fixed and is retained locally at both authorized ends.

\subsection{Privacy Guard Module}

\begin{figure}[t]
    \centering \vspace{-0.3cm}
    \includegraphics[width=0.44\textwidth]{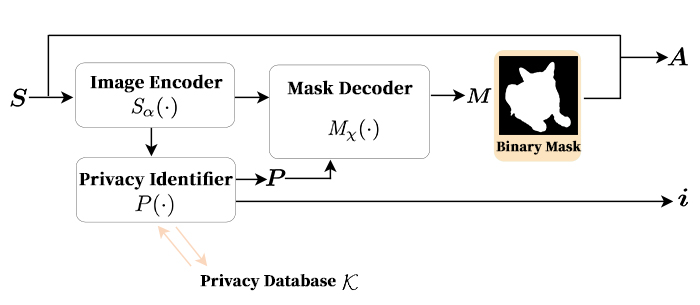} \vspace{-0.2cm}
    \caption{\label{fig_pg}Privacy Guard Module} \vspace{-0.4cm}
\end{figure}

The privacy guard module comprises three sections: image encoder $S_\alpha(\cdot)$, privacy identifier $P(\cdot)$, and mask decoder $M_\chi(\cdot)$, as illustrated in Fig. \ref{fig_pg}. 
Image encoder $S_{\alpha}(\cdot)$ transforms the input image $\boldsymbol{S}$ into a feature map, which is then fed into both privacy identifier and  mask decoder. 
In the privacy identifier, each local patch of the feature map is compared with the stored feature vector $\boldsymbol{f}_k$ for all $k$ using cosine similarity. 
If the similarity with any privacy entity $k$ exceeds a predefined  threshold $\Gamma$, the patch is marked as sensitive. 
The matched entity index is recorded in $\boldsymbol{i} = k$, and the patch location is recorded in $\boldsymbol{P}$. This process is repeated across all patches to generate the complete outputs $\boldsymbol{i}$ and $\boldsymbol{P}$.

Given the feature map of the input image and the identified  location of the sensitive patch $\boldsymbol{P}$, the mask decoder $\textit{M}_{\chi}(\cdot)$ aggregates the spatial information to generate a precise binary mask. 
The generation process can be expressed as:
\begin{equation}
\boldsymbol{M} = M_{\chi}\left( S_{\alpha}(\boldsymbol{S}), \boldsymbol{P} \right)   \in \{ 0,1\}^{H \times W},
\end{equation}
\noindent
where $\boldsymbol{M}_{h,w}=1$ indicates that the pixel at $(h,w)$ belongs to a privacy region while $\boldsymbol{M}_{h,w}=0$ denotes a non-privacy area.  
Therefore, the privacy-removed image can be given as:
\begin{equation}
\boldsymbol{A} =(\mathbf{1}- \boldsymbol{M})\odot\boldsymbol{S}, 
\end{equation}
\noindent
where $\odot$ denotes the element-wise product, and $\boldsymbol{A} $ is the resulting image with sensitive content removed. 
The procedure of the privacy guard module is summarized in Algorithm \ref{alg1}, and the overall training will be provided in  Algorithm \ref{alg3}.

\begin{algorithm}[t]
\caption{Privacy Guard Algorithm} \label{alg1}
\begin{algorithmic}[1]\small
\State \textbf{Initialization:} Load the trained model ${S}_{\alpha}(\cdot)$ and ${M}_{\chi}(\cdot)$, and construct the privacy database $\mathcal{K}$.
\State \textbf{Input:} Image $\boldsymbol{S}$ \\
Process ${S}_{\alpha}(\boldsymbol{S})$ to get the feature map
\If{Privacy information is detected}
    \State Retrieve privacy index $\boldsymbol{i}$ and record the location in $\boldsymbol{P}$
    \State Generate the privacy mask: $\boldsymbol{M} \gets M_{\chi}\left( S_{\alpha}(\boldsymbol{S}), \boldsymbol{p} \right)$
    \State Apply privacy mask: $\boldsymbol{A} \gets (\mathbf{1}- \boldsymbol{M})\odot\boldsymbol{S}$
\Else
    \State $\boldsymbol{A} \gets \boldsymbol{S}$
\EndIf

\State \textbf{Output:} $\boldsymbol{A}$, $\boldsymbol{i}$
%\begin{itemize}
 %   \item If privacy is detected: the non-privacy image $\boldsymbol{a}$ and privacy index $\boldsymbol{i}$.
  %  \item Otherwise: the original image $\boldsymbol{s}$ only.
%\end{itemize}
\end{algorithmic}
\end{algorithm}\normalsize

\subsection{Source-Channel Encoder and Decoder}

To enable robust wireless transmission, the privacy-removed image $\boldsymbol{A}$ and its associated semantic index $\boldsymbol{i}$ are jointly encoded into a unified bitstream, which is interpreted as a one-hot message class\cite{oshea2018introduction}. 
For a bitstream of length $b$, there exist $2^b$ distinct messages, each represented by a unique class label. 
These messages are then transformed into a channel input vector $\boldsymbol{x}$ by the source-channel encoder,  subject to the SNR threshold $\tau_{\text{thre}}$ and the transmit power constraint $P_{\text{max}}$, to ensure the compliance with the constrains in (\ref{equ_opt}). 
After the signal goes through the channel, the source-channel decoder then infers the corresponding class label from the received message $\boldsymbol{y}$. 
Finally, the reconstructed image $\boldsymbol{\hat{A}}$ and semantic index $\boldsymbol{\hat{i}}$ are obtained. The source-channel encoder and decoder are jointly trained in an end-to-end manner, following the procedure described in Algorithm \ref{alg3}.

\subsection{Privacy Recovery Module}

\begin{figure}[t] %\vspace{-0.1cm}
    \centering
    \includegraphics[width=0.48\textwidth]{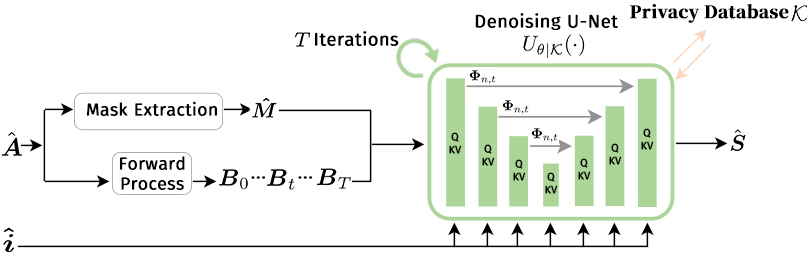}\vspace{-0.2cm}
    \caption{\label{fig_pr}Privacy Recovery Module} \vspace{-0.6cm}
\end{figure}

The privacy recovery module incorporates  a VLM-based diffusion framework to regenerate the removed sensitive content, as shown in Fig. \ref{fig_pr}. 
First, a binary mask \(\hat{\boldsymbol{M}} \in \{0,1\}^{H \times W} \) is obtained by detecting the color differences along the boundaries of the privacy region in $\boldsymbol{\hat{A}}$.

In parallel, the received image $\boldsymbol{\hat{A}}$ goes through a forward diffusion process, where Gaussian noise $\boldsymbol{\epsilon} \sim \mathcal{N}(\boldsymbol{0}, \mathbf{I})$ is gradually added over $T$ timesteps onto the image. 
At each step $t=0,\cdots, T$, the noisy latent variable is computed as:  
\begin{equation}
  \boldsymbol{B}_t   =   \sqrt{\bar{\alpha}}_t \cdot\boldsymbol{\hat{A}} + \sqrt{1 - \bar{\alpha}_t} \cdot \boldsymbol{\epsilon}  ,
\end{equation}
where  $ \bar{\alpha}_t \ge 0 $ is time-dependent hyperparameter that decreases with $t$. At $t=0$, $\bar{\alpha}_0$ equals to 1, so $\boldsymbol{{B}}_{0}=\boldsymbol{\hat{A}}$ contains no noise, and as $t$ increases, noise is progressively added, producing a sequence of latent variables $\{ \boldsymbol{B}_t \}_{t=0}^T$.

Next, a denoising U-Net $U_{\theta|\mathcal{K}}(\cdot)$ is employed to reconstruct  the removed privacy content. 
The model parameters $\theta$ are fine-tuned on the privacy database $\mathcal{K}$ via transfer learning, as summarized in Algorithm \ref{alg2}. 
The inputs to $U_{\theta|\mathcal{K}}(\cdot)$ include the binary mask $\boldsymbol{\hat{M}}$ and the noisy latent sequence $\{ \boldsymbol{B}_t \}_{t=0}^T$  generated from the forward diffusion process. Meanwhile, the privacy-entity index $\hat{\boldsymbol{i}}$  provides conditional vision-language context to guide the image generation and ensure semantic consistency within the masked region. 

The denoising process starts from timestep $t=T$, where the input $\boldsymbol{B}_{T}$ is processed by the U-Net to produce the intermediate output: $\boldsymbol{C}_{T-1} = U_{\theta|\mathcal{K}}(\boldsymbol{B}_{T}| \boldsymbol{\hat{i}})$, where  $\boldsymbol{\hat{i}}$ serves as vision-language condition. 
The  generated content is then masked to retain only the privacy region, while the non-privacy background is filled using pixels from the corresponding noisy image $\boldsymbol{B}_{T-1}$, i.e.,
\begin{equation}
    \boldsymbol{D}_{T-1} =  \underbrace{\boldsymbol{C}_{T-1} \odot \boldsymbol{\hat{M}}}_{\text{\small Masked-region generation}} + \underbrace{\boldsymbol{B}_{T-1} \odot (\mathbf{1}- \boldsymbol{\hat{M}})}_{\text{\small Background retention}},
\end{equation}
which serves as the input for the next time step $t=T-1$. 
This process is repeated iteratively for  $t=T-1,\cdots,1$ via 
\begin{equation}
    \boldsymbol{D}_{t-1} = U_{\theta|\mathcal{K}}(\boldsymbol{D}_t| \boldsymbol{\hat{i}}) \odot \boldsymbol{\hat{M}}  +  \boldsymbol{B}_{t-1} \odot (\mathbf{1}- \boldsymbol{\hat{M}}). 
\end{equation} 

The finial output $\boldsymbol{D}_{0}$ corresponds to the regenerated image $\boldsymbol{\hat{S}}$, with the removed privacy content semantically reconstructed by the receiver.

\begin{algorithm}[t]
\caption{\label{alg2}Transfer Learning Based on Privacy Database $\mathcal{K}$}
\begin{algorithmic}[1]\small
\Statex \textbf{Initialization:} Load the pre-trained model $U_{\theta}(\cdot)$
\State \textbf{Input:} $\mathcal{K} = \left\{ i_k, \mathcal{S}_k = \{ \tilde{\boldsymbol{S}}_{k,m} \}_{m=1}^{M}, \boldsymbol{f}_k \right\}_{k=1,\dots,N}$

\For{$k = 1, \cdots, N$}
    \State $i_k, \mathcal{S}_k \gets \mathcal{K}$
    \For{$m = 1, \cdots, M$}
        \State $\tilde{\boldsymbol{S}}_{k,m} \gets \mathcal{S}_k$
        \State $\{ \boldsymbol{B}_{t,k,m} \}_{t=0}^T \gets$ forward diffusion on $\tilde{\boldsymbol{S}}_{k,m}$ %Generate noisy samples: 
        \For{$t = T, \cdots, 1$}
            \State $\boldsymbol{C}_{t-1,k,m} \gets U_{\theta}(\boldsymbol{B}_{t,k,m}| \boldsymbol{i}_k)$
            \State Compute  $\mathcal{L}_{t,k,m} = \| \boldsymbol{C}_{t-1,k,m} - \boldsymbol{B}_{t-1,k,m} \|^2$
            \State Update $\theta$ using gradient descent on loss $\mathcal{L}_{t,k,m}$
        \EndFor
    \EndFor
\EndFor

\State \textbf{Output:} $U_{\theta \mid \mathcal{K}}(\cdot)$

\end{algorithmic}
\end{algorithm}\normalsize

%%%%%%%%%%%%%%%%%%%%%%%%%%%%%
Furthermore, to enable conditional generation of the removed content based on the index $\hat{\boldsymbol{i}}$, the U-Net $U_{\theta|\mathcal{K}}(\cdot)$ is augmented  with a cross-attention mechanism \cite{vaswani2017attention}.   
In each cross-attention layer, the query $\mathbf{Q}$ is computed from intermediate feature maps of the U-Net, while the key $\mathbf{K}$ and value $\mathbf{V}$ are derived from the text embedding $\boldsymbol{\hat{i}}$: 
\begin{align}
\text{Attention}(\mathbf{Q}, \mathbf{K}, \mathbf{V}) &= 
\mathrm{softmax}\left( \frac{\mathbf{Q} \mathbf{K}^{T}}{\sqrt{d}} \right) \mathbf{V}, 
\end{align}
\begin{align}
\mathbf{Q} = \mathbf{W}_Q \cdot \boldsymbol{\Phi}_{n,t}, \quad 
\mathbf{K} = \mathbf{W}_K \cdot \boldsymbol{\hat{i}}, \quad 
\mathbf{V} = \mathbf{W}_V \cdot \boldsymbol{\hat{i}}, 
\end{align}
\noindent
where $\boldsymbol{\Phi}_{n,t}$ denotes the output of $n$-th intermediate layer within the U-Net at timestep $t$,  $\mathbf{W}_Q$, $\mathbf{W}_K$, $\mathbf{W}_V$ are learnable projection matrices, and $d$ is the scaling factor. The overall training procedure is provided in  Algorithm \ref{alg3}.

\begin{algorithm}[t]
\caption{\label{alg3}Training Semantic Communication Framework}
\begin{algorithmic}[1]\small

\State \textbf{Input}: Image set $\{ \boldsymbol{S}\}$, privacy database $\mathcal{K} = \left\{  \boldsymbol{i}_k, {\mathcal{S}}_k, \boldsymbol{f}_k  \right\}_{k=1}^{N}$, channel parameter $\tau_{\text{thre}}$ and $P_{\text{max}}$

\State \textbf{Transmitter}:
\State \hspace{1em} $(\boldsymbol{A}, \boldsymbol{i}) \gets G_\gamma(\boldsymbol{S}| \mathcal{K})$
\State \hspace{1em} $\boldsymbol{x} \gets E_\eta(\boldsymbol{A}, \boldsymbol{i} |\tau_{\text{thre}}, P_{\text{max}})$
\State \hspace{1em} Transmit $\boldsymbol{x}$ over the channel

\State \textbf{Authorized Receiver}:
\State \hspace{1em} Receive $\boldsymbol{y}$
\State \hspace{1em} $(\hat{\boldsymbol{A}}, \hat{\boldsymbol{i}}) \gets D_\psi(\boldsymbol{y}|\tau_{\text{thre}}, P_{\text{max}})$
\State \hspace{1em} $\hat{\boldsymbol{S}} \gets R_\lambda(\hat{\boldsymbol{A}}, \hat{\boldsymbol{i}} | \mathcal{K})$

\State \textbf{Benign Eavesdropper} (for training purpose only):
\State \hspace{1em} Receive $\boldsymbol{y}_e$
\State \hspace{1em} $\hat{\boldsymbol{S}}_e \gets R_e(D_e(\boldsymbol{y}_e))$

\State \textbf{Loss Computation}:
\State \hspace{1em} Compute distortion loss: $\mathcal{L}_a$
\State \hspace{1em} Evaluate privacy leakage: $f(\hat{\boldsymbol{S}}_e)$ via $G_\gamma$
\State \hspace{1em} Total loss: $\mathcal{L}_{\text{total}} = \mathcal{L}_a + f(\hat{\boldsymbol{S}}_e)$

\State \textbf{Optimization:}
\State \hspace{1em} Update $\gamma, \eta, \psi, \lambda$ via gradient descent on $\mathcal{L}_{\text{total}}$

\State \textbf{Output:} $G_\gamma(\cdot)$, $E_\eta(\cdot)$, $D_\psi(\cdot)$, $R_\lambda(\cdot)$

\end{algorithmic}
\end{algorithm}\normalsize

\section{Simulation Results and Analysis}\label{simulation}

\begin{figure*}[t]
    \centering
    \includegraphics[width=0.675\textwidth]{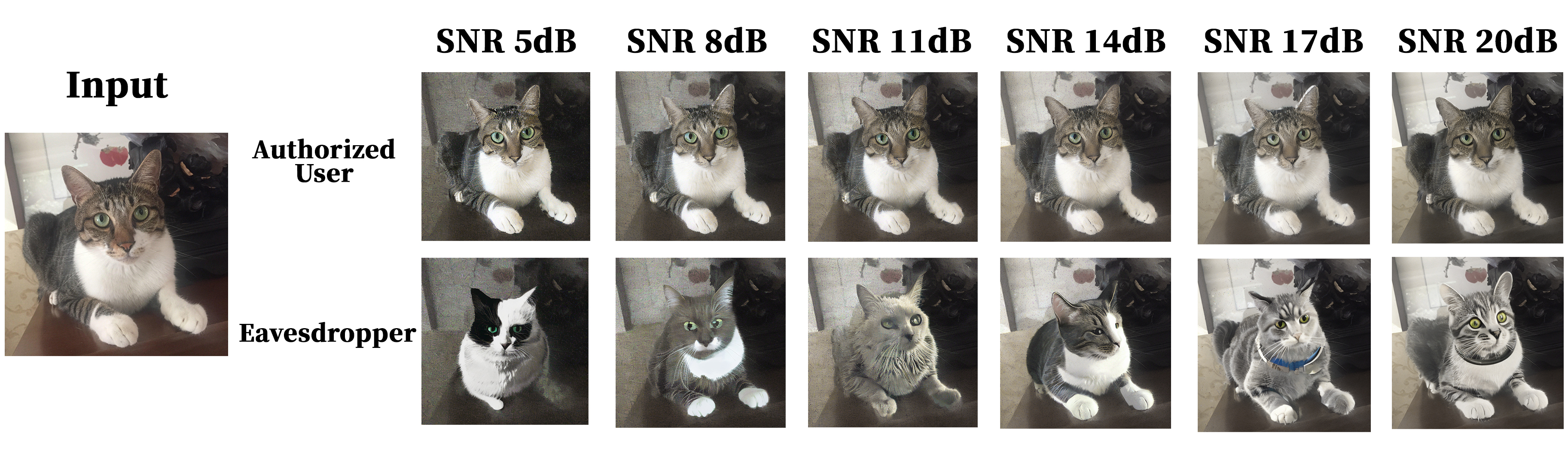} \vspace{-0.2cm}
    \caption{Reconstructed image at the authorized receiver and the eavesdropper under different levels of SNR. }
    \label{fig:result1}
\end{figure*}

In our simulations, a real-world dataset is used to evaluate the performance of the proposed privacy-preserving semantic  framework. 
The dataset comprises $13,536$ images for $518$ individual cats \cite{lin2018catdataset}, captured in diverse natural scenes using standard digital cameras and smartphones. 
The diversity in background, lighting, and pose makes the dataset well-suited for testing real-world applicability.  
For the privacy guard module, we adopt the pretrained  segment anything model (SAM) \cite{kirillov2023segment}, which employs a vision transformer-based image encoder to convert input images into dense feature maps. These feature maps support flexible image segmentation conditional on various input prompts. 
The source-channel encoder and decoder are implemented using an autoencoder framework, trained under different AWGN channel conditions with SNR ranging from $5$ to $20$ dB. 
For the privacy recovery module, we fine-tune stable diffusion model \cite{rombach2022high} using the privacy dataset to reconstruct the removed private content, conditioned on text embeddings. 
For the eavesdropper,   a similar decoder architecture and the same pre-trained stable diffusion model are used, but with no access to the privacy dataset. After decoding the signal, the eavesdropper attempts to recover the removed private content using only the received data and its local generative model.

\begin{figure}[t] \vspace{-0.0cm}
    \centering
    \includegraphics[width=6.5cm]{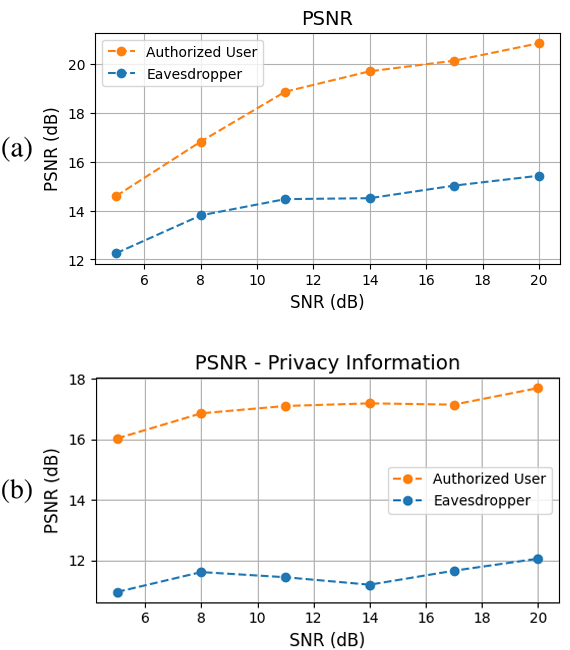}\vspace{-0.1cm}
    \caption{ (a) PSNR of the original and reconstructed images  (b) PSNR of the original and reconstructed privacy content,  at the receiver and the eavesdropper given different SNRs.}
    \label{fig:psnr}    \vspace{-0.4cm}
\end{figure}

\begin{figure*}[t]
	\begin{center} \vspace{-0.0cm}
		\begin{subfigure}{.32\textwidth}
			\centering
			\includegraphics[width=6cm, height=5cm]{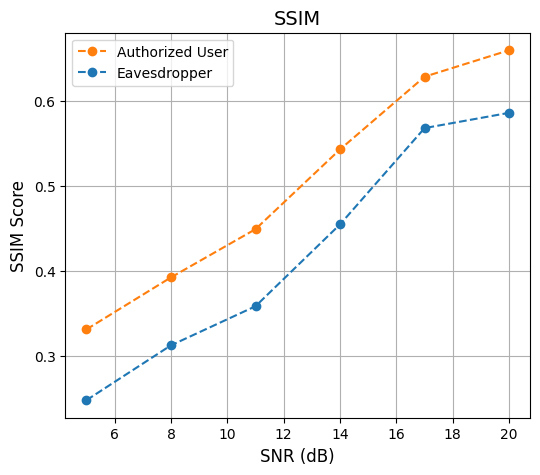}%\vspace{-0.2cm}
			\caption{\label{fig_ssim} }
		\end{subfigure}
		\begin{subfigure}{.32\textwidth}
			\centering
			\includegraphics[width=6cm]{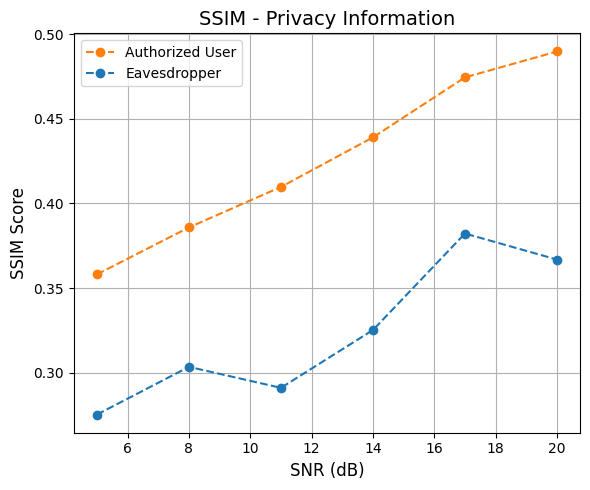}%\vspace{-0.2cm}
			\caption{\label{fig_ssim_m} }
		\end{subfigure}
		\begin{subfigure}{.32\textwidth}
			\centering
			\includegraphics[width=6cm]{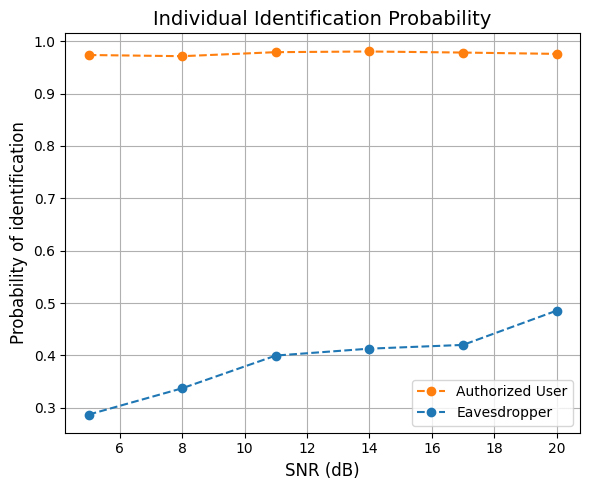}%\vspace{-0.2cm}
			\caption{\label{fig_id}  }
		\end{subfigure}
		\vspace{-0.0cm}
		\caption{\small{\label{fig2} (a) SSIM of the original and reconstructed images at the receiver and the eavesdropper. (b)  SSIM of the original and reconstructed privacy content at the receiver and the eavesdropper. (c)  Probability of identity recognition at the authorized receiver and the eavesdropper. }
		}
	\end{center}
	\vspace{-0.2cm}
\end{figure*}

Fig. \ref{fig:result1} shows the reconstructed images generated by the authorized user and the eavesdropper across different SNR levels. 
An example of input image is shown on the left, with the first row presenting the authorized user's reconstructions and the second row showing the eavesdropper’s outputs.   
As the channel condition improves with higher SNR, the reconstruction quality increases for both parties.  
However, only the authorized user can semantically  recover the cat's identity, while the attacker produces   inconsistent outputs. %distorted or

To evaluate the quality of the reconstructed images, we compare the output of the authorized receiver and eavesdropper, using peak signal-to-noise ratio (PSNR) as metric. 
PSNR measures the pixel-level fidelity, with higher values indicating greater similarity. The PSNR between the original image and the reconstructed image in dB is defined as:
\begin{equation}\small
\text{PSNR}(\boldsymbol{S},\hat{\boldsymbol{S}}) = 10 \log_{10} \left( \frac{\text{MaxValue}^2}{\text{MSE}(\boldsymbol{S},\hat{\boldsymbol{S}})} \right)  ,
\end{equation}\normalsize
where MaxValue is the maximum pixel value, and MSE denotes the mean squared error between the original and reconstructed images. 
As shown in Fig. \ref{fig:psnr}a, image reconstruction quality improves for both the authorized receiver and the eavesdropper as the received SNR increases, due to the enhanced quality of the privacy-removed image $\hat{\boldsymbol{A}}$.  
The authorized receiver, with the prior information from the privacy dataset, consistently achieves PSNR scores at least $2$ dB higher. Moreover, the improvement is more significant for the authorized receiver, as a more accurate semantic index $\hat{\boldsymbol{i}}$ strengthens the construction of the privacy attention maps $\boldsymbol{K}$ and $\boldsymbol{V}$, which further improves image generation quality.

To evaluate the reconstruction of private content, we compute the PSNR  within the privacy region of the reconstructed images. 
As show in Fig. \ref{fig:psnr}b, the authorized receiver with access to the privacy database achieves a PSNR more than $5$ dB higher than the eavesdropper. 
However, as the channel SNR increases, both parties  show only marginal PSNR improvements. 
This is because the masked region is transmitted as an information-less area, thus it has limited contributions to the image reconstruction. 
%As a result, improved transmission quality in this region has limited impact on image recovery.

In addition to PSNR, we use the structural similarity index measure (SSIM) \cite{wang2004image} to evaluate the perceptual quality of reconstructed images as follows:
\begin{equation} \footnotesize
\text{SSIM}  
 = \left( \frac{2\mu_s \mu_r + c_1}{\mu_s^2 + \mu_{{r}}^2 + c_1} \right)^{\alpha_1} 
   \left( \frac{2\sigma_s \sigma_{{r}} + c_2}{\sigma_s^2 + \sigma_{{r}}^2 + c_2} \right)^{\alpha_2} \left( \frac{\sigma_{s{r}} + c_3}{\sigma_s \sigma_{{r}} + c_3} \right)^{\alpha_3},
\end{equation}\normalsize
where $\mu_s$ and $\mu_{{r}}$ are the means, $\sigma_s$ and $\sigma_{{r}}$ are the standard deviations, and $\sigma_{s{r}}$ is the cross-covariance of the original and reconstructed images.  
Constant $c_1, c_2, c_3$ stabilize the computation when the denominator is close to zero. The coefficients $\alpha_1$, $\alpha_2$, $\alpha_3$ weight the contributions of luminance, contrast, and structure components.
The value of SSIM ranges from $0$ to $1$, with higher values indicating better preservation of the structural fidelity in the reconstructed image.
Unlike PSNR which emphasizes pixel-wise differences, SSIM measures the structural similarity between two images, and thus, offers a close alignment with human visual perception. 
%（Deleted on August 12）
% By comparing local patterns of pixel intensities, SSIM effectively quantifies the perceptual degradation that may not be reflected in PSNR. 
%Mathematically, SSIM is defined as: %the product of three components: luminance similarity, contrast similarity, and structural similarity, as %Each component is raised to a tunable exponent to control its contribution to the final score. s
As shown in Figs. \ref{fig_ssim} and \ref{fig_ssim_m}, SSIM improves for both the authorized user and the eavesdropper as the channel SNR increases. 
However, the authorized user consistently achieves higher structural fidelity, which is approximately $10\%$ higher over the entire image and $25\%$ higher within the privacy region, due to access to the privacy database and the aid of text embeddings.

Finally, to test the identity preservation,  the reconstructed images from both the authorized receiver and the eavesdropper are fed back into the privacy guard module to determine whether the private entity can be correctly identified.  
For a fair evaluation,  all test images are previously unseen by the semantic modules and are used exclusively to measure identification performance.  
As shown in Fig. \ref{fig_id}, the authorized receiver achieves near $100\%$ identification accuracy, while the eavesdropper succeeds in less than $50\%$ of the cases.  
These results show that the proposed privacy-preserving semantic  framework effectively protects sensitive identities against a strong eavesdropper in the majority of cases.

\section{Conclusion}\label{conclusion}

In this paper, we have proposed a VLM-based framework for privacy-preserving semantic communications that prevents sensitive content leakage during wireless transmission. 
By aligning the transmitter and receiver through a shared privacy dataset, the system identifies and masks private content before transmission and reconstructs the relevant content at the receiver using a generative model guided by shared semantic priors. 
Simulation results have shown that the proposed framework preserves semantic fidelity for authorized users, effectively limits information recovery by eavesdroppers, and generalizes well to unseen image processing tasks.

\section*{Acknowledgment}
This work was supported in part by the U.S. National Science Foundation under Grant ECCS-2434054. %Any opinions, findings, and conclusions or recommendations expressed in this material are those of the authors and do not necessarily reflect the views of the National Science Foundation.

\bibliographystyle{IEEEtran}
\bibliography{references}

\end{document}